\documentclass[prb,aps,twocolumn]{revtex4}
\usepackage{graphicx,amssymb,amsmath,color,psfrag}
\usepackage{amsthm}
\usepackage{amsfonts}
\usepackage{algorithmic}
\usepackage{enumerate}
\usepackage{latexsym}

\begin{document}
\newcommand{\dt}{\Delta\tau}
\newcommand{\al}{\alpha}
\newcommand{\ep}{\varepsilon}
\newcommand{\ave}[1]{\langle #1\rangle}
\newcommand{\have}[1]{\langle #1\rangle_{\{s\}}}
\newcommand{\bave}[1]{\big\langle #1\big\rangle}
\newcommand{\Bave}[1]{\Big\langle #1\Big\rangle}
\newcommand{\dave}[1]{\langle\langle #1\rangle\rangle}
\newcommand{\bigdave}[1]{\big\langle\big\langle #1\big\rangle\big\rangle}
\newcommand{\Bigdave}[1]{\Big\langle\Big\langle #1\Big\rangle\Big\rangle}
\newcommand{\braket}[2]{\langle #1|#2\rangle}
\newcommand{\up}{\uparrow}
\newcommand{\dn}{\downarrow}
\newcommand{\bb}{\mathsf{B}}
\newcommand{\ctr}{{\text{\Large${\mathcal T}r$}}}
\newcommand{\sctr}{{\mathcal{T}}\!r \,}
\newcommand{\btr}{\underset{\{s\}}{\text{\Large\rm Tr}}}
\newcommand{\lvec}[1]{\mathbf{#1}}
\newcommand{\gt}{\tilde{g}}
\newcommand{\ggt}{\tilde{G}}
\newcommand{\jpsj}{J.\ Phys.\ Soc.\ Japan\ }

\title{Behavior of a Magnetic Impurity in Graphene in the Presence of a Vacancy}
\author{F. M. Hu$,^{1}$ J. E. Gubernatis$,^2$ Hai-Qing Lin$,^{3}$ Yan-Chao Li$,^{3}$ R. M. Nieminen$^1$}
\affiliation{$^1$COMP/Department of Applied Physics, Aalto University School of Science, P.O. Box 11000, FI-00076 Aalto, Espoo, Finland\\
$^{2}$ Theoretical Division, Los Alamos National Laboratory, Los Alamos, New Mexico 87545, USA\\
$^{3}$ Beijing Computational Science Research Center, Beijing 100084, China}

\begin{abstract}
With quantum Monte Carlo methods, we investigate the consequences of placing a magnetic adatom adjacent to a vacancy in a graphene sheet. We find that instead of the adatom properties depending on the energy of the adatom orbital, as in a single impurity problem, they develop a dependence on the energy of the split localized state associated with the single vacancy problem.  Shifting the chemical potential through this experimentally more accessible  energy scale reveals novel behavior in the spectral density, magnetic susceptibility, and the correlations of the adatom spin and charge with those of the conduction electrons. In general, the behavior of the adatom in the presence of a vacancy differs significantly from its behavior in the absence of a vacancy.
\end{abstract}

\pacs{73.22.Pr, 71.55.Jv, 75.30.Hx}
\date{\today}
\maketitle

\begin{figure}[t]
\begin{center}
\includegraphics[scale=0.45, bb=-75 630 316 785]{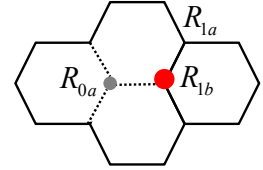}
\end{center}
\caption{(color online). (a) The vacancy-adatom system. The vacancy is at the site $R_{0a}$ and the adatom (red ball) is at its nearest-neighbor site $R_{1b}$. The next-nearest-neighbor site is $R_{1a}$.}\label{Fig:struc}
\end{figure}

\section{Introduction}
The interest in the consequences of defects in graphene has been mushrooming. \cite{Ugeda10,Chen11,Tao11} Graphene's linear density of states imparts unusual properties to the behavior of its defects which in turn generates unusual physical properties for the graphene sheets. The interest is especially high in controlling these defect-induced properties, particularly for those cases where the adatom develops a magnetic moment.


Usually vacancies or magnetic adatoms in graphene are individually studied, and they have been found to exhibit unusual properties. For example, a vacancy leads to the formation of a sharp doubly degenerate localized state sitting at the junction of the upper and lower bands. This localized state is quite robust to the addition of many vacancies. In a nearest-neighbor, tight-binging particle-hole symmetric model of graphene, the amplitude of the resonant state wavefunction is zero on all lattices sites that are on the same sublattice of the vacancy. \cite{Pereira06}

A magnetic adatom on graphene behaves quite differently from the classic Anderson impurity physics coupling to a constant density of states. Generally, with a linear density of states, a Kondo effect does not occur, \cite{Withoff90} and such observables as magnetic susceptibilities and spectral densities differ markedly from mean-field and constant density of state renormalization group theories. As a function of model parameters, the adatom magnetic moment in graphene can cross-over from having free moment Curie-like temperature dependence to having a paramagnetic-like temperature independent one. The spectral densities lack a central peak. Instead, they have two peaks, one in the upper band and one in the lower, separated by roughly $U$, the Coulomb interaction associated with the impurity orbital.\cite{Cornaglia09,Hu11}

In this work, we report the consequences of an adjacent vacancy-adatom pair. We find that when paired these defects exhibit new unusual features that differ from either the single-vacancy or the single-adatom case. For example, the robust central resonance of the single vacancy problem disappears, even in the particle-hole symmetric case.  The upper and lower peaks of the single adatom problem are each split with the two outer peaks of the now four separated by roughly $U$, and the positions of the inner two have only a weak dependence on $U$. These peaks are the doubly degenerate localized vacancy states shifted to the upper and lower bands by the presence of the adatom.  The significance of this split is that by changing the chemical potential of this two-defect system, we can turn the adatom magnetic moment on and off by varying it over a energy range set by the localized vacancy state instead of a range set by the location of the adatom impurity orbital. We also find that the two-defect wavefunction  has a small amplitude on sublattice sites the same as that of the vacancy, just as in the  case of a single vacancy.

\section{Model and Methods}
Our starting point is the Anderson impurity model which has a single impurity orbital of energy $\varepsilon_{d}$ and Coulomb repulsion $U$  inhibiting its occupancy by two electrons. \cite{Anderson61} The impurity orbital is coupled to a free-electron conduction band with hybridization of strength $V$. The total Hamiltonian is
\[
H=H_{0}+H_{1}+H_{2}.
\]
As standard, $H_0$ is a tight-binding Hamiltonian which for graphene is
\[
H_{0}=-t\sum_{<ij>,\sigma}[a^{\dag}_{i\sigma}b_{j\sigma}^{}+b^{\dag}_{j\sigma}a_{i\sigma}^{}]-\mu\sum_{<ij>,\sigma}[a^{\dag}_{i\sigma}a_{i\sigma}^{}+b^{\dag}_{i\sigma}b_{i\sigma}^{}]\texttt{,}
\]
where $a^{\dag}_{i\sigma}$ and $b^{\dag}_{i\sigma}$ create an electron with spin $\sigma$ at sites $R_{ia}$ and $R_{ib}$ on the  $A$ and $B$ sub-lattices of graphene's hexagonal structure. In graphene the hopping matrix element $t>0$ and is about 2.8 eV \cite{Rmp}. $\mu$ is chemical potential and in ideal graphene its value is 0. $H_1$ is the impurity Hamiltonian
\[
H_{1}=\sum_{\sigma}(\varepsilon_{d}-\mu)d^{\dag}_{\sigma}d_{\sigma}^{}+Ud^{\dag}_{\uparrow}d_{\uparrow}^{}d^{\dag}_{\downarrow}d_{\downarrow}^{}\texttt{.}
\]
where $d^{\dag}_{\sigma}$ creates an electron with spin $\sigma$ at the impurity orbital. Finally, $H_{2}$ describes the hybridization between the adatom impurity  and one of graphene's carbon atoms. As shown in Fig.~\ref{Fig:struc}, the tight-binding portion of the Hamiltonian is perturbed by creating a vacancy at site $R_{0a}$.  In what follows we place the adatom either on the opposite sublattice at $R_{1b}$ or on the same sublattice at $R_{1a}$. For an adatom on site $R_{1b}$,
\[
H_{2}=V\sum_{\sigma}[b^{\dag}_{1\sigma}d_{\sigma}^{}+d^{\dag}_{\sigma}b_{1\sigma}^{}]\texttt{.}
\]

Our principal computational tool is the single-impurity quantum Monte Carlo (Hirsch-Fye) algorithm for computing the thermodynamic properties of the adatom, and the method of Bayesian statistical inference for computing its spectral density.
The Hirsch-Fye algorithm \cite{hirsch86} naturally returns the imaginary-time Green's function $G_d(\tau)= \sum_\sigma  {G_{d\sigma}  \left( \tau  \right)} $ of the impurity. With this Green's function, we can easily compute for the adatom orbital such basic quantities as the expected values of the total charge $n_d$, the square of the spin $m_d^2$, and the doubly occupancy $n_{d\uparrow}n_{d\downarrow}$. We note that
\begin{equation}\label{eq:moment}
m_d^2=n_d-2n_{d\uparrow}n_{d\downarrow}
\end{equation}
A non-zero value of $m_d^2$ indicates the formation of a moment on the adatom orbital. The closer this value is to one the more fully developed is the moment. We also calculate the static adatom spin susceptibility
\begin{equation}\label{eq:susceptibility}
\chi=\int^{\beta}_{0}d\tau\langle m_{d}(\tau)m_{d}(0)\rangle,
\end{equation}
where $\beta=T^{-1}$, $m_{d}(\tau)=e^{\tau H}m_{d}(0)e^{-\tau H}$.

Computing the imaginary-time Green's function also enables us to compute the spectral density $A(\omega)=\sum_\sigma A_\sigma(\omega)$ by numerically solving \cite{jarrell96}
\[
G_d\left( \tau  \right)  ={\int\limits_{ - \infty }^\infty  {d\omega } } \frac{e^{  -\tau\omega}{A  \left(\omega  \right)}}
{{e^{ - \beta \omega}  + 1}}
\]
The detailed procedure for doing this is presented in Ref.~\onlinecite{jarrell96}. This Bayesian inference procedure is colloquially, but imprecisely,  also called the maximum entropy method.

Invoking the bipartite nature of the lattice , which remains true even with the vacancy and adatom, and using the standard particle-hole transformation on one-sublattice, we can prove that when $\varepsilon_d=-U/2$, $G_d(\tau,\mu)=G_d(-\tau,-\mu)$ and consequently that $A_{\sigma}(\mu,\omega)=A_{\sigma}(-\mu,-\omega)$. These two results in turn imply symmetries with respect the sign of $\mu$ in the various thermodynamic quantities of interest. These symmetries also mean that without loss of generality we can restrict our attention to the behavior of the system when $\mu \le 0$.

We calibrated our quantum Monte Carlo results by performing exact diagonalization studies \cite{Lin93} of an 11-site graphene structure with a vacancy and adatom. This system is half-filled when there are 12 electrons. We started with this filling and then systematically removed electrons, keeping the number of up and down electrons equal when their total number $N$ was even or having one more up than down if their total number was odd. For the same system, we also performed quantum Monte Carlo calculations and found excellent agreement in the computed thermodynamic quantities and correlation functions.

We comment that the exact diagonalization calculation is at zero temperature and in the canonical ensemble. The quantum Monte Carlo simulation is at
a finite temperature and in the grand canonical ensemble. To compare quantum Monte Carlo with exact diagonalization,
we did the simulation at a very low temperature ($\beta=52$) and adjusted the chemical potential so that the average number of electrons closely
matched that of the zero temperature system. \cite{note} 

\begin{figure}[t]
\begin{center}
\includegraphics[scale=0.32, bb=56 65 665 509]{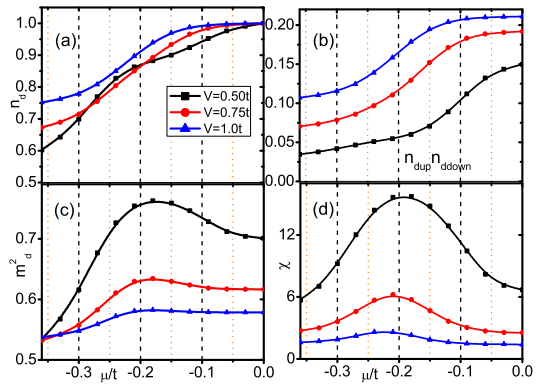}
\end{center}
\caption{ (color online). (a) $n_d$ as a function of $\mu$. (b) $n_{d\text{up}}n_{d\text{down}}$ as a function of  $\mu$. (c) $m_d^2$ as a function $\mu$. (d) $\chi$ as a function of $\mu$. In all plates, $U=0.80t$, $\varepsilon_{d}=-U/2$, and $T^{-1}=32t^{-1}$. } \label{Fig:nomk}
\end{figure}

\section{Results}
\subsection{Basic Thermodynamic Quantities}
In Fig.~\ref{Fig:nomk} we present our study of the behavior of several basic thermodynamic properties as a function of $\mu$ when the adatom is at $R_{1b}$. Here $\varepsilon_{d}=-U/2$, so when $\mu=0$, the system possesses particle-hole symmetry. This symmetry fixes the total charge on the adatom at one. As we see from Fig.~\ref{Fig:nomk}a, by shifting $\mu$ lower, we reduce this charge.  Figure~\ref{Fig:nomk}b shows the expected value of the adatom's double occupancy. The double occupancy is relatively small and becomes smaller as the value of $\mu$ is lowered, but increases as $V$ increases. From these two curves we could predict the values of the $m_d^2$ from Eq.~\ref{eq:moment}. We actually compute it independently.  We show this measure of a local moment in Fig.~\ref{Fig:nomk}c. Despite the general reduction of the total charge and double occupancy as we lower $\mu$, we see that for $V=0.50t$ the moment first increases and then drops in value. For $V=0.75t$ the moments are much smaller with an even smaller initial increase. When $V=1.0t$, the moment is even smaller, and we see no initial enhancement. Fig.~\ref{Fig:nomk}d shows the adatom's spin susceptibility in Eq.~(\ref{eq:susceptibility}). In it is mirrored the behavior of $m_d^2$. The enhancement for $V=0.50t$ is significantly larger than that for the other two cases.

The principal differences among the results in Fig.~\ref{Fig:nomk} are $n_d$ for the $V=0.50t$ case initially decreasing more rapidly and varying differently as $\mu$ varies over the interval (-0.10$t$,-0.20$t$). We note that around $\mu=-0.20t$, $n_d$ for $V=0.50t$ even becomes larger than that of $V=0.75t$. These results point to an anomalous situation, where as we decrease the total charge on the adatom orbital, which must be one or less, we increase the instantaneous net spin of these electrons.

We comment that if were to place the adatom at $R_{1a}$, we would find much smaller enhancements if any occur. We recall that in the absence of an adatom the wave-function of the localized vacancy state is non-zero at $R_{1b}$ but zero at $R_{1a}$. We therefore expect differences in the adatom properties for the two different placements.

\begin{figure}[t]
\begin{center}
\includegraphics[scale=0.45,bb=37 215 474 428]{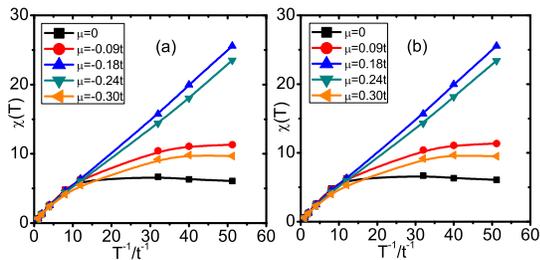}
\end{center}
\caption{(color online). (a) The spin susceptibility $\chi$ with negative $\mu$ as a function of $T^{-1}$ (b) $\chi$ with positive $\mu$. $U=0.80t$, $V=0.50t$ and $\varepsilon_d=-0.40t$.} \label{Fig:KTmu}
\end{figure}

Figure~\ref{Fig:KTmu} shows the adatom's spin susceptibility $\chi$ as a function of inverse temperature for different values of positive and negative $\mu$ for $V=0.50t$ and $\varepsilon_{d}=-U/2$. To numerical accuracy, the results for the different signs of $\mu$ are identical. For $\mu=0$, a particle-hole symmetric case, we see the susceptibility becoming temperature independent at low temperature, even though a fairly well developed moment exists (Fig.~\ref{Fig:nomk}c). This co-occurrence is precedented but unusual. The temperature independence develops because the vacancy and adatom states are well below $\mu$. As we lower $\mu$ a Curie-like temperature dependence develops and then crosses over to a very weak temperature dependence.


\begin{figure}[t]
\begin{center}
\includegraphics[scale=0.40, bb=16 48 611 442]{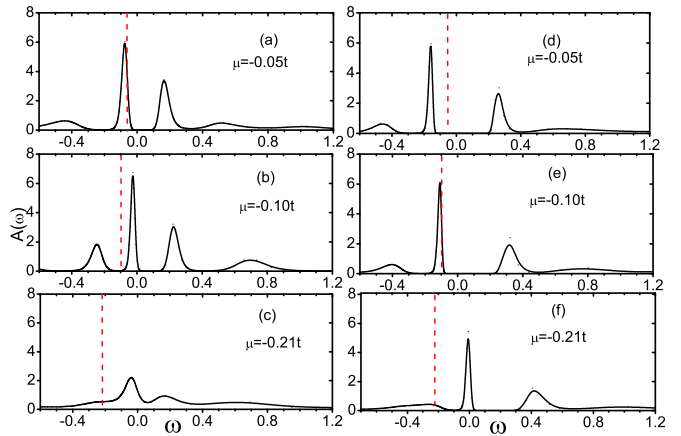}
\end{center}
\caption{ (color online). The spectral density $A(\omega)$ versus $\omega$. (a)-(c) $V=0.50t$ and (d)-(f) $V=1.0t$.} \label{Fig:spec}
\end{figure}

\begin{figure}
\begin{center}
\includegraphics[scale=0.37, bb=74 36 595 434]{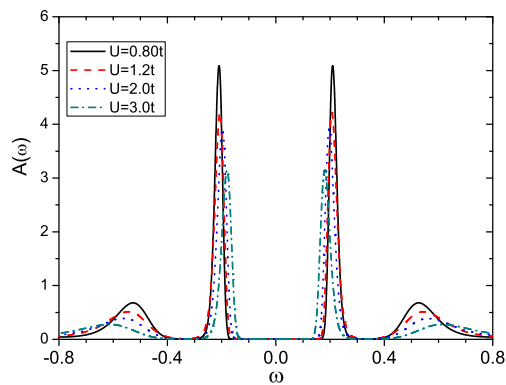}
\end{center}
\caption{ (color online). The spectral density $A(\omega)$ versus $\omega$ with different values of $U$. Here $V=1.0t$ and $\varepsilon_d=-U/2$.} \label{Fig:spec_U}
\end{figure}

\subsection{Spectral Densities}
We also calculated the adatom's spectral densities. For the $V=0.50t$ case we show our results in Fig.~\ref{Fig:spec}a-c for $\mu=-0.05t$, $-0.10t$ and $-0.21t$. The dashed lines mark the location of the chemical potential: $\omega=\mu$. In all sub-figures $U=0.80t$, $\varepsilon_{d}=-U/2=-0.40t$, $T=32t^{-1}$. We clearly see the four-peak structure noted earlier plus the breakdown of particle-hole symmetry as soon as $\mu$ is slightly shifted from zero. Figure~\ref{Fig:spec}d-f for $V=1.0t$ complements the $V=0.50t$ case. In both cases, as we lower $\mu$ the former localized vacancy state peaks are first pushed away from $\omega=0$ and then are drawn back. Consequently, the dynamics of the local moment is initially controlled by the localized vacancy state and then by the adatom impurity state.

We note that when $\mu$ is far from zero, the larger $V$ results give a wider width to $A(\omega)$ peaks associated with the impurity state.
This is not unexpected, but we also note that our procedure for determining the spectral density tends to broaden high frequency features. \cite{jarrell96}

In Fig.~\ref{Fig:spec_U}, we study the spectral densities varying $U$ with the hole-particle symmetry, $\varepsilon_d=-U/2$ and $\mu=0$, and here $T=32t^{-1}$. In bulk transition metal, the value of $U$ is about 5eV to 10eV, \cite{Mahan} and in the case of transition metal atom in carbon-based materials, the value of $U$ is about 2eV to 5eV. \cite{Jacob09,Wehling10B,Jacob10,Wehling11} We change the Coulomb interaction strength in the range (0.8t,3t), which overlaps the above ranges. We see that when U is increased, the outer two peaks become farther from each other while the two inner become closer, and we also note that these changes are small in this range. We expect that the two inner peaks will continue to move toward each other when U is increased.

We comment that the strong central peak characteristic of a vacancy in graphene is absent when the adatom is present. In fact it becomes absent as soon as particle-hole symmetry is absent. The inner peaks in Fig.~\ref{Fig:spec} are these $\mu=0$ localized states shifted by the symmetry breaking and the presence of the adatom. With exact diagonalization we computed the local density of states for several such cases in the absence of the adatom. The only presence of the adatom  perturbs the peak locations slightly.

\begin{figure}[t]
\begin{center}
\includegraphics[scale=0.38, bb=51 60 450 468]{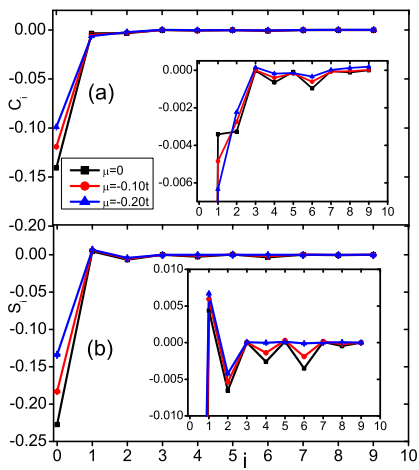}
\end{center}
\caption{ (color online).(a) Charge-charge correlation function $C_i$ (b) Spin-spin correlation function $S_i$. $U=0.80t$, $V=0.50t$ $\varepsilon_d=-0.40t$ and $T^{-1}=32t^{-1}$. The insets show the tails of $C_i$ and $S_i$.}
\label{Fig:corfunc}
\end{figure}

\subsection{Correlation Functions}
The expected physics of the single impurity Anderson model in the absence of a vacancy and the Kondo effect is a double peak structure in $A(\omega)$ where the lower frequency peak is associated with a singly occupied state and the upper peak with an excitation to a doubly occupied state. The single electron in this impurity state is anti-ferromagnetically correlated with the conduction band electrons and fluctuates in its spin polarization so that $\langle n_{d\uparrow}-n_{d\downarrow} \rangle =0$. Fluctuations in charge are relatively weaker.

An extension of the Hirsh-Fye algorithm \cite{Gubernatis87} makes the computation of the the spin-spin $S_i$ and charge-charge $C_i$ correlations between the adatom and conduction electrons possible. We show typical results in Fig.~\ref{Fig:corfunc}. 
$C_0$ and $S_0$ are the on-site correlation functions for an adatom at $R_{1b}$. The locations with even index are positions on the same sublattice as $R_{1b}$ and those with an odd index are positions on the opposite sublattice.  We see that both the spin-spin $S_i$ and charge-charge correlations $C_i$ are relatively short-ranged.

The plots of $S_i$ show that the adatom spin is always ferromagnetically coupled with the conduction electron spin at neighboring sites.  For the single adatom case, this correlation shifts from ferromagnetic to anti-ferromagnetic as we shift $\mu$ below zero. In addition, the correlations are longer-ranged and less damped. We also note that $S_i$ as a function of displacement from the vacancy quickly becomes anti-ferromagnetic when the adatom spin and conduction electron spins are on the same sublattice but represents uncorrelated spins if they are on opposite sublattices. Presumably, as for the single vacancy case, the two-defect wavefunction is very small at the sites in the second case and larger at the sites in the first case. We further note that as we lower $\mu$ the spin correlations become monotonically suppressed, even though over the range of the shift where we are increasing and then decreasing the adatom moment (Fig.~\ref{Fig:nomk}). This suppression is the opposite of the enhancement we observed for the single adatom case. \cite{Hu11}

The charge-charge correlation function $C_i$ is larger if the adatom and conduction electron charges are on the same sublattice. When we shift $\mu$ below zero, its amplitude on these sites also decreases. When we shift $\mu$, the occupancy of the impurity orbital and conduction band shift from half filling so their charge exchange (the possibility of hopping) enhances, and the expected value of the charge decreases (Fig.~\ref{Fig:nomk}),which in turn decreases the size of the correlations. This behavior is also opposite of what we observed for the single adatom case. \cite{Hu11}

\section{Remarks and Conclusions}
We expect the strong features in $A(\omega)$ to appear in the system's density of states. Accordingly, the position of the chemical potential relative to the position of the peaks in $A(\omega)$ can influence the physics of the adatom. When the chemical potential is near or at the location of the shifted localized vacancy state, the physics is that of a localized state. When away, the physics changes. Our results are consistent with this expectation.

Our findings are a result of the impurity orbital now sharing its electrons between two states with two different energies instead of its electrons being associated with a single energy state. Still, these states must share a total charge of one or less, depending on the chemical potential, and both must respect the energy penalty for double occupancy. In turn, these two states hybridize with each other through their hybridization with the conduction electrons and exchange electrons. As we lower the chemical potential and begin to deplete the electron occupancy in the upper localized state, at the same time reducing the double occupancy (Fig.~\ref{Fig:nomk}), we can increase the fluctuations of net non-zero spin and hence increase $m_d^2$. When we leave the localized state behind, we are left with just the impurity state, which now has fewer remaining electrons and strongly suppressed spin and charge fluctuations. In general, as a function of the chemical potential, the presence of the vacancy causes the adatom impurity to behave differently, and in the case of its spin and charge correlations with the conduction electrons, for example, the behavior trends as a function of $\mu$ are opposite to those of the single adatom case.



It is clear that by shifting the chemical potential away from the half-filled particle-hole symmetric case we can have a magnet moment that we can enhance or diminish by changing the chemical potential. We remark that vacancies in graphene are readily generated by irradiation. \cite{Hahn99,Hashimoto04} We also note that organic groups absorbed on the graphene sheet can generate the localized states similar to those induced by vacancies. \cite{Wehling10} So the organic groups also can be regarded as a kind of ideal vacancies in graphene. And finally we remark that scanning tunneling microscopy (STM) opens the possibility for experimentalists to place a variety of magnetic adatoms at precise locations on a graphene sheet. \cite{Eigler90}  In principle, our predictions are testable experimentally.

We have not studied how our predictions change as the separation between the two defects increases, other than to show that the effects are most pronounced if the vacancy and adatom are on different sublattice positions. Clearly, if the system has many of either or both defects, the results will likely change significantly. Many vacancies, for example, fill in graphene's pseudo-gap, removing, on the average, the linear density of states.

\section{Acknowledgement}
This work was supported by Academy of Finland through its Center of Excellence (2006-2011) program. The work of J. E. G. was supported by the U.S. Department of Energy. We acknowledge computational resources from CSC-IT Center for Science Ltd.


\begin{thebibliography}{99}
\bibitem{note} The total number of the electrons $N$ in an exact diagonalization calculation is set before hand, while in quantum Monte Carlo it is unknown before hand but is tunable by adjusting the chemical potential $\mu$. We obtained $N$ using an extension of Hirsch-Fye algorithm . \cite{Gubernatis87}
\end{thebibliography}
\end{document}